\documentclass[aps,pra,reprint,superscriptaddress,showpacs]{revtex4-2}

\usepackage{graphicx}
\usepackage{color}

\begin{document}


\title{Molecular dynamics characterisation of radiosensitising coated gold nanoparticles in aqueous environment}

\author{Alexey V. Verkhovtsev}
\email{verkhovtsev@mbnexplorer.com}
\altaffiliation{On leave from the Ioffe Physical-Technical Institute, Polytekhnicheskaya 26, 194021 St. Petersburg, Russia}
\affiliation{MBN Research Center, Altenh\"oferallee 3, 60438 Frankfurt am Main, Germany}

\author{Adam Nichols}
\affiliation{School of Physical Sciences, Ingram Building, University of Kent, Canterbury, CT2 7NH, UK}

\author{Nigel J. Mason}
\affiliation{School of Physical Sciences, Ingram Building, University of Kent, Canterbury, CT2 7NH, UK}

\author{Andrey V. Solov'yov}
\altaffiliation{On leave from the Ioffe Physical-Technical Institute, Polytekhnicheskaya 26, 194021 St. Petersburg, Russia}
\affiliation{MBN Research Center, Altenh\"oferallee 3, 60438 Frankfurt am Main, Germany}


\begin{abstract}
Functionalised metal nanoparticles (NPs) have been proposed as promising radiosensitising agents for more efficient radiotherapy treatment using photons and ion beams. Radiosensitising properties of NPs may depend on many different parameters (such as size, composition and density) of the metal core, organic coatings and the molecular environment. A systematic exploration of each of these parameters on the atomistic level remains a formidable and costly experimental task but it can be addressed by means of advanced computational modelling. This paper outlines a detailed computational procedure for construction and atomistic-level characterisation of radiosensitising metal NPs in explicit molecular media. The procedure is general and extensible to many different combinations of the core, coating and environment. As an illustrative and experimentally relevant case study we consider nanometre-sized gold NPs coated with thiol-poly(ethylene glycol)-amine molecules of different length and surface density and solvated in water at ambient conditions. Radial distribution of different atoms in the coatings as well as distribution and structural properties of water around the coated NPs
are analysed and linked to radiosensitising properties of the NPs.
It is revealed that the structure of the coating layer on the solvated NPs depends strongly on the surface density of ligands.
At surface densities below $\sim$3 molecules/nm$^{2}$ the coating represents a mixture of different conformation states, whereas elongated ``brush''-like structures are formed at higher densities of ligands.
The water content in denser coatings is significantly lower at distances from 1~nm up to 3~nm from the gold surface depending on the length of ligand molecules.
Such dense and thick coatings may suppress the production of hydroxyl radicals by low-energy electrons emitted from the metal NPs and thus diminish their radiosensitising properties.
\end{abstract}

\maketitle


\section{Introduction}

Functionalised metal nanoparticles (NPs) have become a topic of intense research in view of their potential applications in many different research areas, including electronics, chemistry and nanomedicine \cite{Tsukuda_Hakkinen, Capek_NM_NPs, Yang_2015_ChemRev.115.10410}.
NPs made of noble metals (especially, gold and platinum) and other metals like gadolinium have been proposed as novel and promising agents for more efficient treatment of tumours with ionising radiation \cite{kwatra2013nanoparticles, bergs2015role, yamada2015therapeutic, Haume_2016_CancerNano.7.8}.
Such NPs have the capacity to enhance the biological damage induced by energetic photon and ion-beam irradiation, i.e. to act as radiosensitisers \cite{porcel2010platinum, butterworth2012physical, li2016let, Kuncic_Lacombe_2018_PMB.63.02TR01, Verkhovtsev_2015_PRL.114.063401}. 
When accumulated in the tumour region, these NPs can locally enhance the radiation damage of tumour cells
relative to normal tissue, thereby increasing the efficiency of radiation therapy.

The radiosensitising effect of metal NPs is commonly attributed to strong irradiation-induced emission of secondary electrons \cite{Verkhovtsev_2015_PRL.114.063401, xiao2011role, Verkhovtsev_2015_JPCC.119.11000}.
Secondary electrons induce radiolysis of the surrounding molecular medium (consisting, to a large extent, of water)
and thus increase the yield of hydroxyl radicals and other reactive oxygen species which may produce damage to tumour cells \cite{sicard2014new,gilles2014gold}.
Exploration of the nanoscale mechanisms underlying radiosensitisation by NPs is a subject of current efforts of
numerous research teams worldwide \cite{Nano-IBCT_Springer, Schuemann_PMB_NPs_roadmap, Kempson_2020_IJMS.21.2879}.

Metal NPs for radiotherapies and other biomedical applications are usually synthesised with an organic coating
in order to reduce toxicity, improve stability under physiological conditions and target specific biological sites \cite{Ghosh_2008_AdvDrugDelivRev.60.1307, Kim_2013_AccChemRes.46.681,  NMNP_2019_ChemCommun.55.6964}.
A large number of possible coating molecules, including acids, polymers, sugars and proteins,
form a vast landscape of core--coating combinations.
As a result, the optimal choice in terms of the maximal radiosensitising effect is not known beforehand.
This is partially due to a lack of understanding of the key fundamental processes that take place upon
irradiation of such complex systems.
It has also been shown that the presence of a coating on metal NPs can lower the radiosensitisation potential
of the latter \cite{xiao2011role, gilles2014gold}.
A reduced amount of water near the metal surface may inhibit the production of hydroxyl radicals
\cite{gilles2018quantification, haume2018transport}, although exact mechanisms are not yet understood.
This suggests that dense and thick organic coatings may drastically reduce radiosensitising effects of metal NPs
by preventing water from reaching the surface of the NPs.

A vast number of parameters can be varied to optimise the radiosensitising properties of NPs
(such as the size, shape and composition of a metal core; thickness, composition and density of a coating;
composition and density of a molecular medium) making it a formidable task to systematically explore each
of them experimentally.
It is therefore desirable to develop a computational and theoretical framework for a quantitative atomistic-level description
of the structural properties of coated metal NPs in biologically relevant environments, irradiation-induced effects
(such as electron emission and production of reactive species) as well as the kinetics of irradiation-induced
chemical reactions in the vicinity of NPs.
Such descriptions based on the computational modelling approach may reduce the experimental costs for optimisation,
control and functionalisation of NPs in biomedical applications and thus may facilitate significant progress in the field.

This paper presents a detailed atomistic approach for the computational modelling and structural characterisation of coated metal NPs in molecular media by means of all-atom molecular dynamics (MD) simulations using two advanced software packages,
MBN Explorer \cite{solov2012mesobionano} and MBN Studio \cite{sushko2019modeling}.
Characterisation of the structure of experimentally relevant NPs in realistic environments is a prerequisite for the quantitative description of their radiosensitisation properties.
As an illustrative case study we consider nanometre-sized gold NPs functionalised with thiol-poly(ethylene glycol)-amine (S--(CH$_2$)$_2$--PEG$_n$--NH$_2$), one of frequently used coating materials in radiobiological experiments
involving metal NPs \cite{Grellet_2017_PLOSOne, Zhu_2015_JNanobiotechnol.13.67, Tzelepi_2019_NanoscaleAdv.1.807}.

Gold NPs have previously demonstrated their potential regarding the localisation of radiation-induced
effects \cite{hainfeld2004use, yamada2015therapeutic}.
However, bare gold NPs tend to aggregate under physiological conditions and to be eliminated
quickly from the bloodstream \cite{alexis2008factors}; these effects are reduced through the use of
poly(ethylene glycol) (PEG) coatings \cite{kim2009multimodal}.
Gold NPs coated with PEG-based ligands have therefore been considered for therapeutic and imaging applications \cite{Jokerst_2011_Nanomed.6.715} because of their high stability, biocompatibility, low cytotoxicity,
increased blood circulation time as well as antibacterial activity \cite{Reznickova_2019_ColloidsSurfA.560.26, otsuka2003pegylated}.

\begin{sloppypar}
Recent experiments showed that nanometre-sized gold NPs coated with thiol-PEG-amine
or with a mixture of PEG and galactose sugar molecules are selectively toxic for cancer cells and
hence hold potential as radiosensitisers \cite{Grellet_2017_PLOSOne, Zhu_2015_JNanobiotechnol.13.67}.
It was also demonstrated that shorter PEG molecules and a higher density of sugar molecules can increase
the efficiency of NP uptake by cells, thus revealing the importance of coating structure modulation
for the more efficient NP design \cite{Ding_2014_Part2SystCharact.31.347}.
\end{sloppypar}

A large amount of theoretical and computational work on modelling of coated (also commonly referred to
as ligand-protected) metal NPs and nanoclusters has been performed over the past decade \cite{Tsukuda_Hakkinen, Jin_2016_ChemRev.116.10346, Walter_2008_PNAS.105.9157, Kyrychenko_2015_PCCP.17.12648, pohjolainen2016unified, Haume2016,
Kyrychenko_2017_PCCP.19.8742, Malola_2021_NatComm.12.2197}.
In a recent study \cite{Chan_2020_JPCL.11.2717} the stability and structural properties of gold NPs
(containing $49-79$ atoms) functionalised with PEG$_2$ molecules (made of two EG monomers)
were explored by means of density functional theory (DFT) calculations with dispersion corrections.
In Ref.~\cite{lin2017pegylation} structural and dynamic properties of gold NPs coated with a mixture
of alkane and alkane-PEG molecules were studied by means of coarse-grained MD simulations as functions
of PEG density and the length of PEG molecules.
However, little attention has been paid so far to the atomistic analysis of the impact of structure
and composition of NPs on their radiosensitising properties.

In an earlier study \cite{Haume2016} the structure of a 135-atom gold NP (of $\sim$1.6~nm in diameter)
coated with 32, 48 and 60 thiol-PEG$_5$-amine molecules and solvated in water was explored by means of classical MD simulations.
In the follow-up study \cite{haume2018transport} a novel theoretical and computational approach
to analyse electron emission from coated metallic NPs irradiated with ions was presented.
Two systems, namely Au$_{135}$ coated with 32 and 60 thiol-PEG$_5$-amine molecules, were considered
as illustrative examples.
This was the first atomistic-level study that accounted for the realistic structure of the coated metal NPs
as well as for relevant low-energy and many-body phenomena in the formation and transport of secondary electrons
and free radicals arising after irradiation of the NPs.

In this paper the molecular details of experimentally relevant coatings are systematically explored by means of
all-atom MD simulations using the MBN Explorer \cite{solov2012mesobionano} and MBN Studio \cite{sushko2019modeling} software.
The structure of 1.6-nm gold NPs coated with thiol-PEG$_n$-amine ($n = 2-11$) molecules and solvated
in water is systematically characterised as a function of both ligand surface density and the length of ligands.
The chosen size of the metal core is close to the size of experimentally studied NPs, $(1.9 \pm 0.8)$~nm
\cite{Grellet_2017_PLOSOne, grellet2018optimization}.
The number of ligands attached to the gold core is varied from 8 to 60 for each ligand length,
which corresponds to the ligand surface density of 1 to 7.5 molecules per nm$^2$.
This range of ligand densities has been considered in a vast majority of experimental studies with gold NPs \cite{Smith_2017_Analyst.142.11}.
The systems are chosen according to experimental observations that gold NPs with a size of less than
ten nanometres are typically coated with PEG chains containing less than ten monomers \cite{lin2017pegylation, Ghosh_2008_AdvDrugDelivRev.60.1307, Kim_2013_AccChemRes.46.681, Ghosh_2010_JACS.132.2642}.

It is demonstrated that structure of the coating layer of the solvated NPs depends strongly on the surface density of ligands.
At densities below about 3~nm$^{-2}$ the coating represents a mixture of different conformation states,
whereas elongated ``brush''-like structures are formed predominantly at higher densities of ligands.

The simulations predict the formation of a dense layer of water molecules around the metal core at ligand densities below 2~nm$^{-2}$.
This effect occurs due to strong van der Waals interaction between gold atoms and oxygen atoms of water molecules.
A 10--30\% increase of water density over ambient in close proximity to the gold surface may enhance the production
of hydroxyl radical due to low-energy electrons emitted from metallic NPs \cite{Verkhovtsev_2015_PRL.114.063401, Verkhovtsev_2015_JPCC.119.11000}.
For denser coatings, the density of water in the coating layer is lower than the ambient water density at distances
from 1~nm up to 3~nm from the gold surface depending on the length of ligand molecules.
This may decrease the probability of radical production \cite{haume2018transport} and thus diminish the radiosensitising properties of such NPs.


\section{Computational Methodology}
\label{sec:Methodology}

\begin{sloppypar}
This section describes the step-by-step atomistic approach towards computational modelling of coated metal NPs in molecular environments.
The computational protocol is general and can be used to construct NPs of different size and compositions of the metal core
and the coating.
The composition and properties of a molecular medium (e.g. density or temperature) can also be varied easily.
As a case study the protocol has been applied to construct and characterise gold NPs coated with thiol-PEG-amine
molecules of different length. The systems were solvated in water at ambient density and 300~K temperature.
\end{sloppypar}

The simulations were performed using MBN Explorer \cite{solov2012mesobionano}, a software package for
advanced multiscale modelling of complex molecular structure and dynamics.
MBN Studio \cite{sushko2019modeling}, a multi-task toolkit and a dedicated graphical user interface for MBN Explorer,
was used to construct the systems, prepare input files and analyse simulation outputs.

\subsection{Construction of Free NPs}

\subsubsection{Metal Core}
\label{sec:Methods_metal_core}

As a first step, a gold NP with the diameter of 1.6~nm was cut from an ideal gold crystal by means of the modeller tool of MBN Studio \cite{MBN_Tutorials_2017}.
The geometry of the NP containing 135 atoms was optimised using the velocity quenching algorithm and then annealed in the following way.
The NP was first heated up to 500~K over a time period of 5~ns with a heating rate of 0.1~K/ps;
then, it was maintained at 500~K for 2~ns using the Lanvegin thermostat and cooled down to 0~K over 5~ns in a mirror procedure to heating.
Relaxation at 500~K temperature enables rearrangement of surface atoms while keeping the core region
close to its initial crystalline structure.

It is known that sulphur atoms in thiolated gold clusters and self-assembled monolayers on gold surfaces
form covalent bonds with gold adatoms or gold atoms located at surface \cite{Hakkinen_2012_NatChem.4.443}.
In order to enable binding of ligands only to surface Au atoms and to avoid penetration of the ligands into the NP
two types of gold atoms corresponding to the core and the surface regions have been defined.
55 atoms located in the core region (defined as a sphere of 6~\AA~radius) have the coordination number corresponding to the fcc structure type; the remaining 80 surface atoms located within a spherical shell of 2~Angstroms are characterised by a reduced coordination number.

Gold atoms within the core and surface regions interact via the many-body Gupta potential \cite{Gupta_1983_PRB.23.6265};
parameters of this interaction were taken from Ref.~\cite{cleri1993tight}.
The interaction between the core and surface gold atoms was described using the Lennard-Jones potential
employing the parameters $\varepsilon = 2.168$~eV and $r_0 = 2.946$~\AA, taken from Ref.~\cite{geada2018insight}.
In the cited paper a polarizable Lennard-Jones potential for metallic gold was introduced and validated against
experimental data on lattice parameters, surface energy, and hydration energy with water.

\subsubsection{Ligand Molecules: Case of thiol-PEG-amine}

\begin{figure}[t!]
\centering
\includegraphics[width=1.0\linewidth]{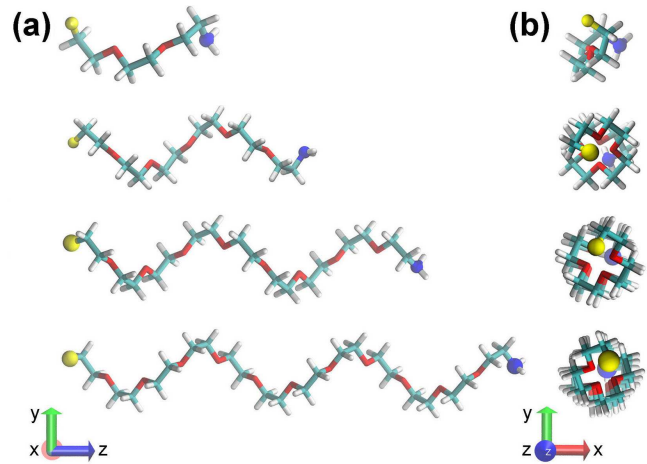}
\caption{Geometries of four thiol-PEG$_n$-amine ($n = 2, 5, 8, 11$) molecules considered in this study: (a)~side view and (b) view along the main symmetry axis. Sulphur and nitrogen atoms are shown by yellow and blue colours, respectively.}
\label{fig:Au135_Au_AuL}
\end{figure}

\begin{sloppypar}
Initial geometries of the thiol-PEG-amine ligands were constructed based on the geometries of isolated PEG molecules.
Coordinates of four PEG molecules with the different number of monomers were taken from the Ligand Expo database \footnote{\lowercase{http://ligand-depot.rutgers.edu/index.html}}.
These molecules, especially those with a larger number of monomers, are folded into helical structures
as shown in Fig.~\ref{fig:Au135_Au_AuL}.
To enable binding of ligands to the metal core each PEG molecule was functionalised with a thiol (SH) group.
The hydrogen atom from the thiol group was then removed and the partial charge sitting on the removed atom ($+0.145|e|$ according to our DFT calculations described below) was equally redistributed over the remaining molecule to keep it electrically neutral.
\end{sloppypar}

In radiosensitisation experiments with gold NPs thiolated PEG molecules are commonly augmented
on the other end with an amino (NH$_2$) group
\cite{Haume_2016_CancerNano.7.8, Grellet_2017_PLOSOne, Zhu_2015_JNanobiotechnol.13.67, Tzelepi_2019_NanoscaleAdv.1.807}.
This modification promotes a more efficient uptake of the NPs by cells through the cellular membrane \cite{Verma_2010_Small.6.12}.
In order to reproduce structures of the experimentally studied systems the remaining terminal oxygen
atom was replaced with a nitrogen atom and an extra hydrogen atom was added.
The resulting thiol-PEG$_n$-amine ($n = 2, 5, 8, 11$) molecules are shown in Fig.~\ref{fig:Au135_Au_AuL}
and their properties are summarised in Table~\ref{tab:PEG_summary}.
Thiol-PEG-amine ligands of such size (with molecular weight of about $200-600$~g/mol) correspond to
the low-mass region of commercially available products for NP synthesis and have been studied in
a series of recent experiments \cite{Grellet_2017_PLOSOne, grellet2018optimization, Tzelepi_2019_NanoscaleAdv.1.807}.

\begin{table*}[!htb]
    \centering
    \caption{Specification of four thiol-PEG-amine ligand molecules considered in this study. 3-symbol identifiers in brackets denote IDs of the original PEG$_n$ molecules in the LigandExpo database. $N_{\rm at}$ is the number of atoms in each ligand molecule; their chemical formulas and molecular weights are listed in the last two columns.}
    \begin{tabular}{p{2.7cm}p{3.7cm}p{1.0cm}p{2.5cm}p{3.0cm}}
    \hline
        Original molecule  &  Ligand molecule  &  $N_{\rm at}$  &  Formula  &  Mol. weight (g/mol) \\
    \hline
        PEG$_3$ (PGE)    &  S-(CH$_2$)$_2$-PEG$_2$-NH$_2$     &   24  &  C$_6$H$_{14}$O$_2$NS         &   164.3   \\
        PEG$_6$ (P6G)    &  S-(CH$_2$)$_2$-PEG$_5$-NH$_2$     &   45  &  C$_{12}$H$_{26}$O$_5$NS      &   296.4   \\
        PEG$_9$ (2PE)    &  S-(CH$_2$)$_2$-PEG$_8$-NH$_2$     &   66  &  C$_{18}$H$_{38}$O$_{8}$NS    &   428.6   \\
        PEG$_{12}$ (12P) &  S-(CH$_2$)$_2$-PEG$_{11}$-NH$_2$  &   87  &  C$_{24}$H$_{50}$O$_{11}$NS   &   560.7   \\
    \hline
    \end{tabular}
    \label{tab:PEG_summary}
\end{table*}

\begin{table}[t!]
    \centering
    \caption{Values of the selected bond lengths $d_{\rm eq}$ and angles $\theta_{\rm eq}$ used as parameters of the CHARMM force field in the simulations presented in this study. These parameters have been generated using the SwissParam web-service \cite{SwissParam_paper} as described in the text. Parameters for the Au--S--C and S--Au--S angular interactions are taken from the indicated references.
    Reference values obtained from DFT calculations carried out in this work and in Refs.~\cite{Jarvi2011development, Bae_2013_JPCA.117.10438, pohjolainen2016unified} are listed for comparison. }
    \begin{tabular}{p{1.3cm}|p{1.7cm}p{1.0cm}|p{4.0cm}}
    \hline
                  & \multicolumn{2}{c}{This study}   & Reference values \\
    \hline
                  &  CHARMM   &   DFT       &  \\
    \hline
        Bond      & \multicolumn{3}{c}{$d_{\rm eq}$~(\AA)} \\
    \hline
        C--C      &   1.508       &  1.513  &     \\
        C--N      &   1.451       &  1.459  &     \\
        C--O      &   1.418       &  1.408  &     \\
        S--C      &   1.801       &  1.834  &  1.78 \cite{Jarvi2011development}, 1.84 \cite{Bae_2013_JPCA.117.10438}    \\
       Au--S      &   2.290       &  2.290  &  2.29 \cite{Jarvi2011development}, 2.28 \cite{Bae_2013_JPCA.117.10438}, 2.33 \cite{pohjolainen2016unified} \\
    \hline
            Angle      & \multicolumn{3}{c}{$\theta_{\rm eq}$~(deg.)} \\
    \hline
        S--C--C   &   109.5       &  109.4  & 105.0 \cite{Jarvi2011development}    \\
        S--C--H   &   116.6       &  109.6  &     \\
        N--C--C   &   108.3       &  111.0  &     \\
        C--C--O   &   108.1       &  108.0  &     \\
        C--O--C   &   106.9       &  112.8  &     \\
       Au--S--C   &   106.8 \cite{pohjolainen2016unified}       &  102.0   & 100.0 \cite{Jarvi2011development}, 105.5 \cite{Bae_2013_JPCA.117.10438}     \\
       S--Au--S   &   172.4 \cite{pohjolainen2016unified}       &          & 160.0 \cite{Jarvi2011development}     \\
    \hline
    \end{tabular}
    \label{tab:CHARMM_param}
\end{table}

\begin{sloppypar}
Interatomic interactions within the ligands and with surrounding water molecules were described using
the CHARMM molecular mechanics force field \cite{MacKerell_1998_JPCB.102.3586}.
The topology of the molecules and parameters of the force field were determined using the SwissParam web-service \cite{SwissParam_paper} \footnote{\lowercase{https://www.swissparam.ch/}}.
Parameters obtained were validated through a series of DFT calculations performed using Gaussian 09 software \cite{Gaussian09}.
In these calculations structures of the pristine and thiolated PEG molecules containing 2 to 5 monomers
as well as the molecules bound to a gold atom were optimised.
The M062X exchange-correlation functional and a 6-31+G(d,p) basis set were employed in the calculations for the pristine and thiolated molecules. For the complexes with a gold atom, a mixed LanL2DZ/6-31+G(d,p) basis set was used, wherein the former set described the metal atom and the latter was applied to C, O, H, N and S atoms.  The difference in partial charges in the different molecules was found to be very small so that it was neglected in further analysis. The optimised geometries were used to perform potential energy scans for selected covalent bonds and angles.
\end{sloppypar}

The parameters of the CHARMM force field used in the simulations are in good agreement with the results of our DFT calculations and the results of other calculations reported in literature \cite{Jarvi2011development, Bae_2013_JPCA.117.10438, pohjolainen2016unified}.
A number of selected equilibrium bond lengths and angles are listed in Table~\ref{tab:CHARMM_param}.
Note that the CHARMM parameters utilized in this study include the bonded and angular interactions involving gold and sulphur atoms. Accounting for these interactions enables the formation of Au--S covalent bonds and S--Au--S staples in which two thiolated ligands are covalently bound to a gold atom located on the surface of a gold NP \cite{pohjolainen2016unified}.

\subsubsection{Construction of Coated Metal NPs}
\label{sec:optimization_NPs_vacuum}

The next step of the computational protocol is devoted to the construction of a coating layer of a given density around the metal NP.
It should be stressed that this study is focused on the site-unspecific attachment of thiol-PEG-amine ligands onto the annealed gold NPs.
In experiments involving radiosensitising gold NPs the size of NPs cannot be controlled precisely and the typical size distribution of NPs is characterised by the standard deviation of about one nanometer \cite{Grellet_2017_PLOSOne, grellet2018optimization}.
The constructed NPs thus represent a prototype of a statistically probable system which is synthesised in experiments involving radiosensitising NPs. The detailed structural analysis of high-symmetry thiolated gold clusters, such as Au$_{102}$(SR)$_{44}$, Au$_{130}$(SR)$_{50}$ or Au$_{144}$(SR)$_{60}$ (where R denotes a ligand molecule) which have been extensively studied in literature \cite{pohjolainen2016unified, Jin_2010_Nanoscale.2.343, Wang_2021_NanoscaleAdv.3.2710}, goes beyond the scope of this study. Nevertheless, structural characteristics of a Au$_{144}$ cluster coated with 60 thiol-PEG$_2$-amine ligands are briefly discussed below in Section~\ref{sec:results_characterisation} in the context of validation of the presented computational approach.

As a first step a spherical layer containing a given number of randomly oriented thiol-PEG-amine molecules
was constructed by means of the modeler plug-in of MBN Studio \cite{sushko2019modeling, MBN_Tutorials_2017}.
The geometry and topology of all molecules comprising the coating layer were recorded in the .pdb and .psf file formats respectively.
8 to 60 ligand molecules of each length were attached to the metal core, corresponding to the surface densities of 1 to 7.5 molecules per nm$^2$.

The covalent interaction between surface gold atoms and sulphur atoms was described using the Morse potential.
The equilibrium distance of 2.29~\AA~between Au and S atoms was determined in our DFT calculations.
The strength of Au--S interaction was set equal to 2.0~eV, which is close to the value of 2.08~eV
reported in Ref.~\citenum{Jarvi2011development}.
The covalent Au--S interaction was truncated at a cutoff distance of 6~\AA.
Parameters for the Au--S--C and S--Au--S angular interactions enabling the formation of sulphur-- gold staples
were taken from Ref.~\citenum{pohjolainen2016unified}.
Parameters of the Lennard-Jones potential describing the non-bonded van der Waals interaction for gold atoms
were also taken from Ref.~\citenum{pohjolainen2016unified}.
The non-bonded interaction was specified for all gold atoms of the NP, including both the core and the surface regions.
Parameters for all other atoms were generated using the SwissParam web-service \cite{SwissParam_paper}.
Parameters of the Lennard-Jones potential for different heteroatomic pairs were then derived using the Lorentz-Berthelot mixing rules,
$\varepsilon_{ij} = \sqrt{\varepsilon_{i} \, \varepsilon_{j}}$ and $r_{0,ij} = (r_{0,i} + r_{0,j})/2$, where the parameters for atoms of a given type are listed in Table~\ref{tab:CHARMM_nonbonded}.

\begin{table}[t]
    \centering
    \begin{tabular}{c|cc}
    \hline
                  &  $\varepsilon$~(kcal/mol)   &   $r_0$~(\AA)  \\
    \hline
        Au        &   5.289        &  2.95   \\
        C         &   0.055        &  4.35   \\  
        N         &   0.200        &  3.70   \\  
        O         &   0.152        &  3.54   \\  
        S         &   0.250        &  3.99   \\  
        H(--C)    &   0.022        &  2.64   \\  
        H(--N)    &   0.046        &  0.45   \\  
        H (water) &   0.046        &  0.45   \\  
        O (water) &   0.152        &  3.54   \\  
    \hline
    \end{tabular}
    \caption{Parameters of the Lennard-Jones potential describing the van der Waals interaction for gold atoms, atoms of the coating and atoms of the surrounding water medium, used in this study. }
    \label{tab:CHARMM_nonbonded}
\end{table}

Attachment of ligand molecules onto the metal core was simulated in a series of short MD runs lasting
10 to 20~ps with the time step of 0.2~fs.
To enable fast and efficient attachment of ligands, the positions of the gold atoms were fixed in space
and the strength of Au--S interaction was set to 100~eV, that is about 50 times higher that the physical Au--S interaction.
The constant-temperature simulations were performed at 1000~K allowing sulphur atoms to move over
the NP surface and eventually find optimal positions.
Due to low symmetry of the metal core after the annealing process and the absence of gold adatoms enabling the site-specific bonding of ligands, only 10--20\% of ligand molecules have formed S--Au--S staples. In most cases, only one ligand was bound to a surface Au atom.
Once the ligands were attached, the Au--S interaction was gradually decreased down to 2.0~eV
(still keeping gold atoms fixed in space); this value was used in all subsequent simulations.

As an additional benchmark of the simulation procedure we have constructed and characterised a Au$_{144}$ cluster coated with 60 thiol-PEG$_2$-amine molecules. The initial geometry of this system was created based on the geometry of the Au$_{144}$(SR)$_{60}$ cluster (where R = phenyl-ethanethiol (PET)) taken from Ref.~\cite{pohjolainen2016unified}. Coordinates of gold and sulphur atoms were adopted from the cited study, and the PET ligands have been replaced with thiol-PEG$_2$-amine molecules.
A series of energy minimisation calculations were then performed with the frozen metal core (including S--Au--S staples) to relax the coating structure.

In the next step the constraints put on gold atoms were lifted, allowing them to move freely.
Energy minimisation calculations for the coated NPs were performed using the velocity quenching algorithm.
Simulations were run for 50,000 steps with a step time of 0.5~fs.

At this point the procedure branched into two directions; all system types (i.e. different ligand lengths
and surface densities) were studied in both cases, with the key difference being the stage at which
the systems were annealed.
In the first instance, coated metal NPs were annealed in vacuum prior to solvation in water, whereas
in the second approach they were solvated first and then annealed together with a whole water box.
The final goal of the whole procedure was to construct physically correct systems.
NPs placed in water were equilibrated at 300~K temperature for sufficiently long times
before final analysis as described in greater detail in
Section~\ref{sec:constuction_solvated_NPs}.

\subsubsection{Annealing of Coated Metal NPs}
\label{sec:annealing_in_vacuum}

At this stage the now coated and optimised systems were annealed in vacuum. The annealing process consisted of heating and cooling stages.

\begin{figure*}[htb!]
    \centering
    \includegraphics[width=0.75\linewidth]{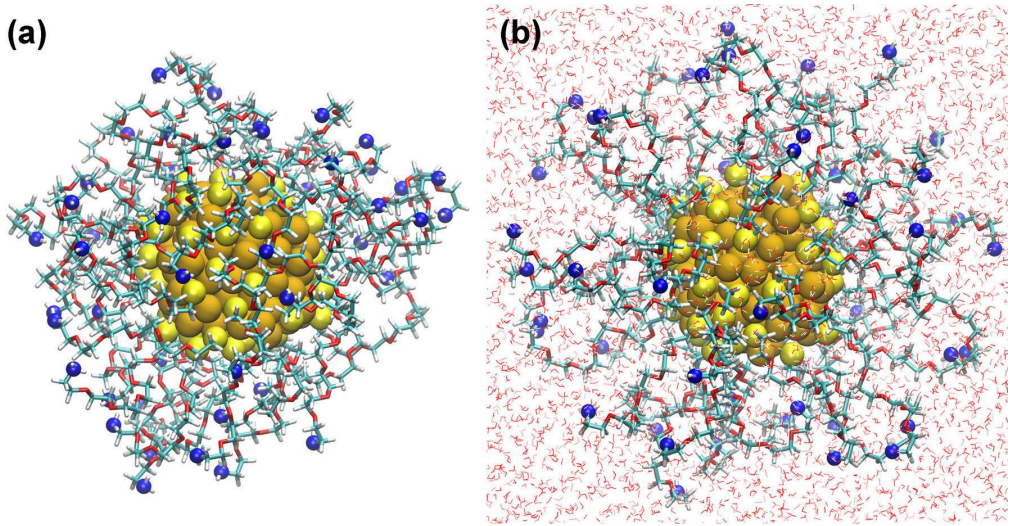}
    \caption{Structure of a 1.6 nm gold NP coated with 48 thiol-PEG$_5$-amine molecules.
    Panel (a): final structure obtained after annealing of the NP in vacuum
    (see Section~\ref{sec:annealing_in_vacuum}).
    Panel (b): final structure obtained after solvation of the NP in water and equilibration of the whole system for 5~ns
    (see Section~\ref{sec:equilibration_water_final}).
    Glossy yellow and blue spheres show sulphur and nitrogen atoms, respectively.}
    \label{fig:AuPEG_vac_vs_wat}
\end{figure*}

In the first simulation the coated gold NPs were placed in a 10~nm length cubic simulation box.
MD simulations were performed for 2~ns with time step 0.1~fs.
A Langevin thermostat was used with 0.5~K/ps temperature increase rate and 0.2 ps damping time.
The thermostat temperature increased from 0~K to 1000~K in the course of the simulation.
In the second simulation the NPs were cooled down to 0~K with the same parameters and the cooling rate of 0.5~K/ps.
Figure~\ref{fig:AuPEG_vac_vs_wat}(a) shows the structure of an annealed Au$_{135}$ NP coated with 48 thiol-PEG$_5$-amine
molecules (corresponding to surface density of 6 molecules per nm$^2$).
The figure illustrates that after the annealing process the ligands are densely curled around the core.

\subsection{Construction of Coated NPs in a Molecular Environment}
\label{sec:constuction_solvated_NPs}

\subsubsection{Introducing NPs to a Solvent}

The thiol-PEG-amine coated gold NPs (either optimised or annealed in vacuum) were solvated in water
by means of MBN Studio \cite{sushko2019modeling}.
The TIP3P water model \cite{TIP3P_model} was used to describe the solvent.
It should be noted that the workflow presented does not rely on any specific water model and the same
procedure can be repeated with an alternative water model if deemed necessary.
The number density of water molecules was set to 33.4~nm$^{-3}$ that corresponds to the accepted number density
of water molecules in a liquid medium at ambient conditions.

Nanoparticles coated with thiol-PEG$_2$-amine and thiol-PEG$_5$-amine molecules were placed in the centre
of a cubic water box of 7~nm length.
For thiol-PEG$_8$-amine and thiol-PEG$_{11}$-amine ligands the side length of the box was increased to 9~nm and 10~nm, respectively.
The size of the simulation box was chosen such that the end of the coating and the simulation box boundaries were
separated by a water layer of $1.5 - 2$~nm.

\subsubsection{Optimising Solvated NPs}

The solvated systems were energy minimised employing the velocity quenching algorithm
and periodic boundary conditions in order to relax the structure and eliminate any overlapping atoms.
The Particle Mesh Ewald (PME) algorithm with a cutoff distance of 10~\AA~was employed to describe
electrostatic interactions.
Non-bonded van der Waals interactions were also truncated at a 10~\AA~distance.
Energy minimisation caused water molecules to disperse into a more stable configuration,
allowing larger integration time steps to be used in subsequent simulations without any numerical artefacts,
thus reducing computation time.

Two consecutive geometry optimisation calculations were performed for each system.
For the first stage of optimisation the simulations were performed for 50,000 steps with a time step of 0.01~fs.
The second simulation used the final configuration of the first one as a starting point and
ran for further 50,000 steps with a time step of 0.1~fs.
Small time steps were employed in optimisation simulations to ensure that the subsequent equilibration simulations
would converge without unforeseen numerical instabilities.

\subsubsection{Final Equilibration of the System}
\label{sec:equilibration_water_final}

In order to create physically correct systems for the subsequent analysis the optimisation
was followed by an equilibration simulation, which was also split into two stages.
A 20-ps long MD simulation was conducted to pre-equilibrate each optimised structure at 300~K temperature.
The Langevin thermostat was employed with damping time of 0.2~ps.
After that a 5-ns long production simulation was conducted for each system with a time step of 2~fs.
Root-mean square displacement of atoms of the coated NPs was analysed to ensure that the systems
have been equilibrated over the 5~ns simulation time.

As mentioned above, we have also probed an alternative approach to construct coated NPs in a solution.
In this case the NPs optimised in vacuum
(see Section~\ref{sec:optimization_NPs_vacuum})
were first solvated in water and then annealed together with the water box.
The systems were heated up to 1000~K at a rate of 0.5~K/ps and then cooled down to 0~K at the same rate.
After annealing the systems were equilibrated at 300~K for 5~ns using the above-described protocol.


\section{Results and Discussion}

\begin{figure*}[t]
\centering
\includegraphics[width=0.9\linewidth]{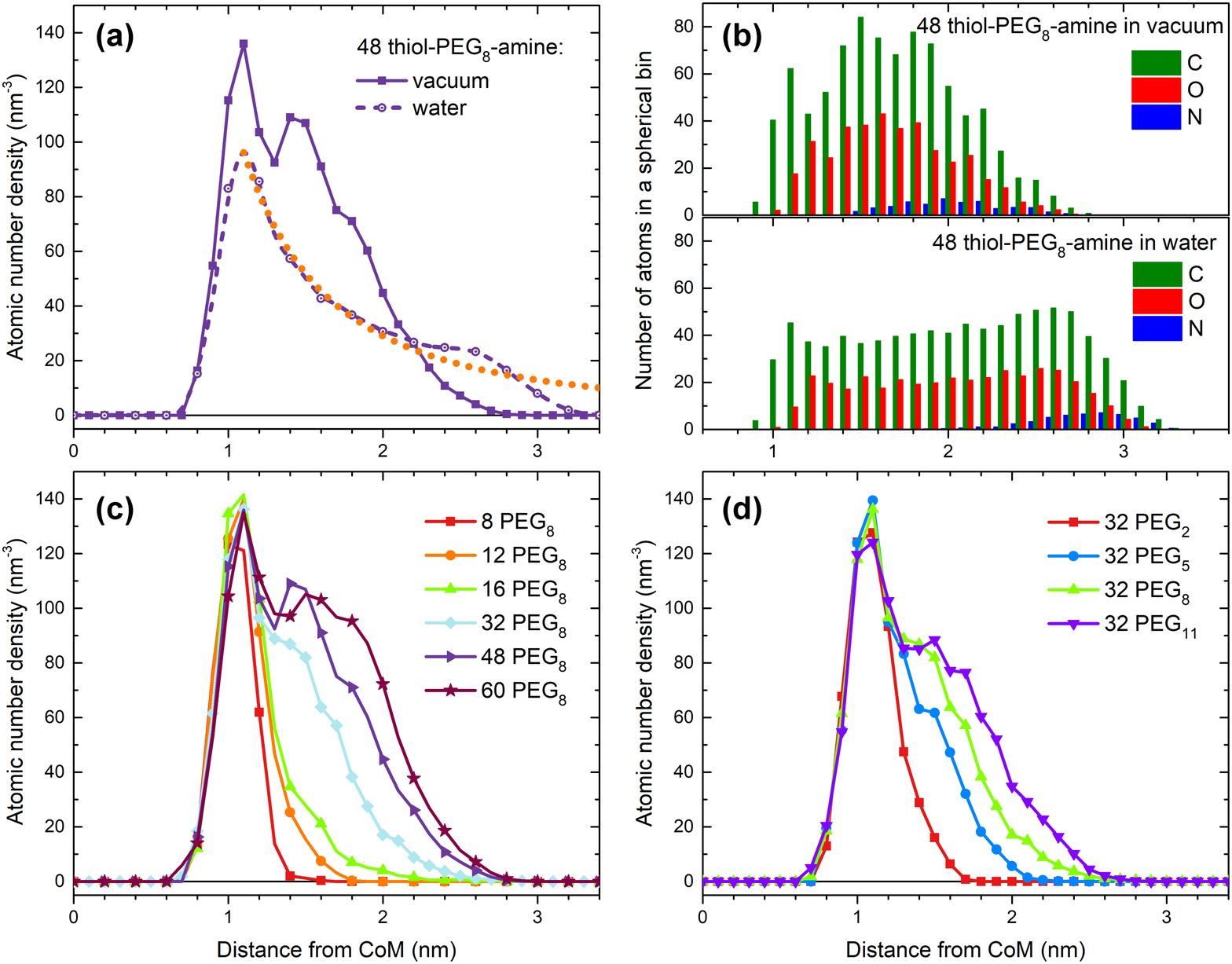}
\caption{Panel~(a): number density $n(r)$ of all atoms in the coating made of 48 thiol-PEG$_8$-amine molecules for NPs annealed in vacuum and solvated in water. The number densities are plotted as functions of radial distance $r$ from the centre of mass (CoM) of the gold core. The dotted orange line shows the $n(r) \propto r^{-2}$ dependence (see the text for details).
Panel~(b): the number of carbon and oxygen atoms of the PEG backbone of thiol-PEG$_8$-amine as well as the number of nitrogen atoms in the terminal NH$_2$ group located in spherical bins of 0.1~nm thickness.
Panels~(c,d): the variation of atomic number density with the number of thiol-PEG$_8$-amine ligands in the coating (c) and with the ligand length for the coatings made of 32 molecules (d).
}
\label{fig:numdens_vac_water}
\end{figure*}

\begin{sloppypar}
We begin the analysis of results by systematically characterising the structure of the thiol-PEG-amine coatings of different length,
both annealed in vacuum and solvated in water.
Particular focus is made on the analysis of radial distribution of different atoms in the coatings.
The possible impact of the type of sulphur--gold bonding (particularly, the formation of S--Au--S staples) on the structural characteristics of gold NPs coated with thiol-PEG-amine is also discussed.
Then, common structural features of the solvated systems
such as a transition from ``mushroom''-like to ``brush''-like structures are elucidated and discussed.
In the ``mushroom'' type of packing ligand molecules attached to the metal core are strongly bent and coiled around the metal surface,
whereas ligands in the ``brush''-like configuration are oriented mainly perpendicular to the surface
\cite{deGennes_1980_Macromol.13.1069, Brittain_2007_JPolymSciA.45.3505, Binder_2012_JPolymSciB.50.1515}.
In the last part of this section the distribution and structural properties of water around the coated NPs as well as links to radiosensitising properties of the NPs are discussed.
\end{sloppypar}

\subsection{Structural Characterisation of Coated NPs}
\label{sec:results_characterisation}

As illustrated in Fig.~\ref{fig:AuPEG_vac_vs_wat} the structure of solvated systems differs significantly from
the structure of free NPs annealed in vacuum.
To quantify this difference Figure~\ref{fig:numdens_vac_water}(a) shows the number density of all atoms in the coating
made of 48 thiol-PEG$_8$-amine molecules, for the solvated NP and the NP annealed in vacuum.
The number densities are plotted as functions of radial distance from the centre of mass (CoM) of the metal core.
In this analysis the whole space with the origin at CoM was divided into spherical bins of 0.1~nm thickness.
The number of atoms of each type located in each bin was then calculated and divided by the volume of each bin.

Figure~\ref{fig:numdens_vac_water}(a) reveals a significant difference between the radial distributions of atoms
in the dense thiol-PEG$_8$-amine coatings annealed in vacuum and equilibrated in water.
The solid line shows the atomic number density distribution for the free NP in vacuum;
the distribution is characterised by two peaks located at 1.1~nm and 1.5~nm from the CoM.
As discussed in detail below each of these peaks is attributed to the formation of a dense layer of ligand molecules.
The first peak located closer to the CoM has been observed for short and low-density coatings, whereas the second peak
emerges with an increase of coating density or ligand length.
The atomic number density distribution for the coating of the solvated NP (dashed line) spans over larger distances from CoM.
The maximum of number density for the coating of the solvated NP is also located at 1.1~nm from the CoM that is $\sim$0.3~nm away from the surface of the metal core.

Figure~\ref{fig:numdens_vac_water}(b) shows the number of carbon and oxygen atoms of the PEG backbone as well as the number of nitrogen atoms in the NH$_2$ group located in spherical bins of 0.1~nm thickness.
In the solvated NP (the lower panel of Fig.~\ref{fig:numdens_vac_water}(b)) the number of carbon and oxygen atoms located in each spherical bin varies little within the distance range from $\sim 1.1$ up to 2.5~nm from CoM.
The relatively uniform distribution of the PEG backbone atoms in the coating layer can explain the decrease of atomic number density for the solvated NP approximately proportional to $r^{-2}$, where $r$ is the distance from CoM (see the dotted orange line in Fig.~\ref{fig:numdens_vac_water}(a)).
This dependence arises due to the $r^2$ increase of the surface area of a spherical bin with the growth of $r$, while the thickness of each bin and the number of atoms located in it stays nearly constant.
A shoulder in the number density distribution located at distances from 2.4~nm to 3~nm (see the dashed line in Fig.~\ref{fig:numdens_vac_water}(a)) corresponds to the position of the terminal NH$_2$ group as indicated by the distribution of nitrogen atoms in the lower panel of Fig.~\ref{fig:numdens_vac_water}(b).

Panels (c) and (d) of Figure~\ref{fig:numdens_vac_water} illustrate the variation of the atomic number density with the number of ligands of particular length and with ligand length for a given number of molecules attached, respectively.
The density profiles for the annealed coatings made of the same number of ligands of different length coincide at distances up to 1.2~nm from CoM (panel~(d)) whereas little variation of the number density is observed within this range of distances as the number of ligands in the coating increases (panel~(c)).
This means that the structure of coatings in the vicinity of the metal core in free NPs does not depend on length and surface density of ligands -- they wrap around the metal core in the ``mushroom'' type of packing \cite{deGennes_1980_Macromol.13.1069, Brittain_2007_JPolymSciA.45.3505, Binder_2012_JPolymSciB.50.1515} and form a dense molecular layer.
Figures~\ref{fig:numdens_vac_water}(c,d) show that as the coating becomes denser as the length of ligand molecules increases, the second peak in the atomic number density distributions emerges at a distance of about 1.5~nm from CoM.
This peak is attributed to the formation of a second dense layer of ligands, which is also seen in Fig.~\ref{fig:AuPEG_vac_vs_wat}(a) for a coating made of 48 thiol-PEG$_5$-amine molecules.

Analysis of the NPs annealed in vacuum reveals that the maximum value of the atomic number density for the coatings
is $\sim$140 nm$^{-3}$,
and this value is almost independent of the number of ligands attached and their length, see Fig.~\ref{fig:numdens_vac_water}(c,d).
The peak number density can be estimated using the following geometrical model.
Ligand molecules shown in Fig.~\ref{fig:Au135_Au_AuL} can be decomposed into structural units each containing a C$_2$H$_4$ fragment bound to an oxygen or a sulphur atom. 
There are three such units in the thiol-PEG$_2$-amine molecule and 12 such units in thiol-PEG$_{11}$-amine.
The linear size of the isolated ligands considered in this study varies from 0.85~nm for thiol-PEG$_2$-amine to 3.3~nm for thiol-PEG$_{11}$-amine, hence one structural unit has the linear size $l_0 \approx 0.28$~nm.
The average cross sectional area of the ligand molecules is $A_{\rm {lig}} = \pi r_0^2$ with $r_0 \approx 0.25$~nm.
Hence the average volume occupied by one such structural unit is equal to $V = \pi r_0^2 \, l_0 \approx 0.055$ nm$^3$.
Let us analyze the ``mushroom'' type of packing where all ligand molecules are tightly packed around the metal core.
A spherical shell of thickness $2r_0 \approx 0.5$~nm surrounding the gold core has a volume of 7.1~nm$^3$ and
thus it can accommodate up to $\sim 130$ structural units in the ``mushroom'' type of packing.
Taking into account that each structural unit contains eight atoms the maximal atomic number density
for the spherical shell of 0.5~nm thickness is $\sim 145$~nm$^{-3}$.
This value corresponds to the peak density in the ``mushroom'' type of packing and it agrees with
the results of the MD simulations shown in Fig.~\ref{fig:numdens_vac_water}(c,d).

This geometrical model can be extended to describe the formation of the second peak in the atomic number density distributions.
The volume occupied by a single thiol-PEG$_8$-amine molecule (with the linear size of 2.5~nm) is about 0.7 nm$^3$;
therefore about 10 such molecules in the form of ``mushroom''-like structures can fit into the aforementioned
spherical shell without overlapping.
For denser coatings containing a larger number of molecules, the molecules would repel each other due to
electrostatic and van der Waals interactions and thus they would stretch away to larger distances from CoM.
This estimate agrees with the results shown in Fig.~\ref{fig:numdens_vac_water}(c), which indicates
broadening of the right tail of the number density distribution for 12 thiol-PEG$_8$-amine molecules
and the formation of a shoulder at distances from 1.4 to 1.8~nm from CoM in the number density distribution for 16 thiol-PEG$_8$-amine.
As the number of attached molecules increases further this shoulder develops into a well-defined peak.

\begin{figure*}[t]
    \centering
    \includegraphics[width=0.75\linewidth]{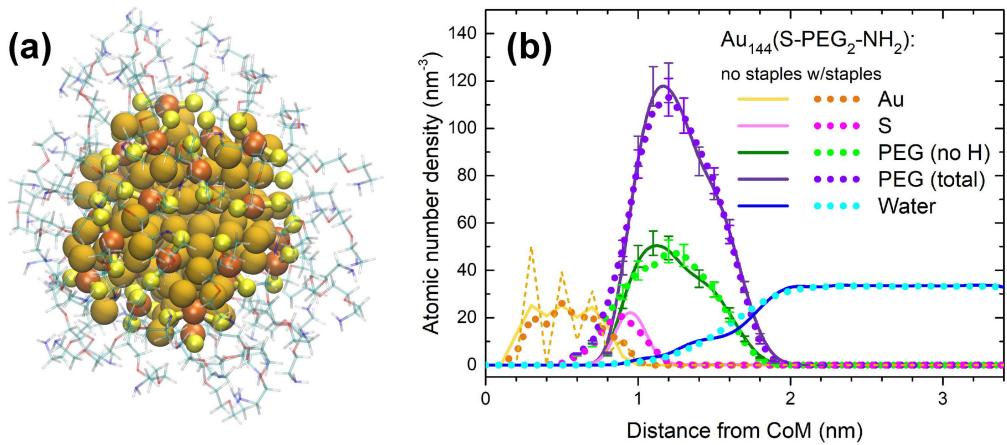}
    \caption{Panel~(a): Structure of a Au$_{144}$(S-PEG$_2$-NH$_2$)$_{60}$ NP where all the ligands are bound to the metal core through the formation of S--Au--S staples. The figure shows the final structure of the NP after annealing in vacuum. Gold and sulphur atoms are highlighted: 30 surface gold atoms which form S--Au--S staples are shown with ochre colour, the remaining gold atoms are shown by matte orange colour, sulphur atoms are shown in yellow. The S--Au--S staples are explicitly shown.
    Panel~(b): number density distributions for gold atoms of the metal core, sulphur atoms, all atoms of the coating made of 60 thiol-PEG$_2$-amine molecules, as well as water molecules surrounding the NPs solvated in the water medium. Solid lines correspond to the number density distributions for the Au$_{144}$(S-PEG$_2$-NH$_2$)$_{60}$ NP without staples. Dotted lines show the number density distributions for the Au$_{144}$(S-PEG$_2$-NH$_2$)$_{60}$ NP with S--Au--S staples. Five independent simulations have been performed for each system; error bars indicate the corresponding standard deviation. The dashed orange line shows the number density distribution for gold atoms in the initial highly-symmetric Au$_{144}$ core.}
    \label{fig:staples}
\end{figure*}

As shown in Fig.~\ref{fig:numdens_vac_water}(a) the maximal value of the atomic number density for the coating of the solvated NP
is about 70\% of the maximal density for the NP in vacuum. This indicates that thiol-PEG-amine ligands in water
are loosely packed compared to the ``mushroom''-like type of packing in the free NP.
The following estimate shows that ligand molecules in water attain more linear ``brush''-like shapes.
The maximal atomic number density for the ``brush''-like packing is determined by the ratio of surface area
of the metal core, $A_{\rm {core}} = 4 \pi R^2$ (where $R = 0.8$~nm is the core radius), to the cross sectional area
per one ligand molecule, $A_{\rm {lig}}$. For the Au$_{135}$ NPs considered in this study the ratio $A_{\rm {core}}/A_{\rm {lig}} \approx 40$.
If the ligands are oriented normally to the metal surface, 40 aforementioned structural units containing eight atoms
would fit into the spherical shell of thickness $l_0 \approx 0.28$~nm.
Knowing the volume of that shell is equal to 3.1~nm$^3$ one estimates the maximal atomic number density equal to $\sim 100$~nm$^{-3}$,
which agrees well with the result shown in Fig.~\ref{fig:numdens_vac_water}(a).
Further analysis of conformation states of the coating molecules in water is presented in the following sections.

As discussed in Section~\ref{sec:Methodology},
during the attachment of ligands to the gold core only one ligand molecule is predominantly bound to a Au atom leading to the formation of Au--S links, while some fraction of ligands form S--Au--S staples through binding of two ligands to one Au atom.
In order to investigate the possible impact of the sulphur--gold bonding on the structural parameters of radiosensitising NPs, structural characterisation of the Au$_{144}$(S-PEG$_2$-NH$_2$)$_{60}$ NP has been carried out.
We have considered NPs in which all 60 ligands are bound to the gold core through the formation of S--Au--S staples (denoted hereafter as Au$_{144}$(S-PEG$_2$-NH$_2$)$_{60}$ with staples) and also NPs created by the site-unspecific attachment of ligands (denoted as Au$_{144}$(S-PEG$_2$-NH$_2$)$_{60}$ without staples), following the procedure outlined in
Section~\ref{sec:Methodology}.

Figure \ref{fig:staples}(a) shows the structure of the Au$_{144}$(S-PEG$_2$-NH$_2$)$_{60}$ cluster with staples after annealing in vacuum.
Figure~\ref{fig:staples}(b) shows a comparison of the number density distributions for the gold atoms of the metal core and atoms of the coatings for Au$_{144}$(S-PEG$_2$-NH$_2$)$_{60}$ without staples (solid lines) and with S--Au--S staples (dotted lines).
Five independent simulations have been performed for each system; error bars indicate the corresponding standard deviation.
The density distributions for atoms of the coating and for water molecules surrounding the solvated NPs are practically the same for the two systems. This indicates that the radiosensitising properties of the coated gold NPs depend very weakly (if at all) on the mechanism of sulphur-gold bonding.
The main difference between the structural characteristics of the two systems is that the number density of sulphur for Au$_{144}$(S-PEG$_2$-NH$_2$)$_{60}$ with staples is shifted by $\sim$0.2~nm towards the CoM (see dotted magenta line) compared to the cluster without staples (pink solid line).
Annealing of the cluster with staples results in a stable but low-symmetry structure where sulphur atoms are distributed over a broader range of distances from CoM compared to the cluster without staples.
Stability of both systems has been confirmed by nanosecond-long MD simulations both in vacuum and in the water medium.
A more detailed structural characterisation of Au$_{144}$(SR)$_{60}$ and other high-symmetry ligand-protected gold clusters, e.g. Au$_{102}$(SR)$_{44}$ or Au$_{130}$(SR)$_{50}$, is an interesting topic which might be addressed in a separate study.

\begin{figure}[t]
    \centering
    \includegraphics[width=0.85\linewidth]{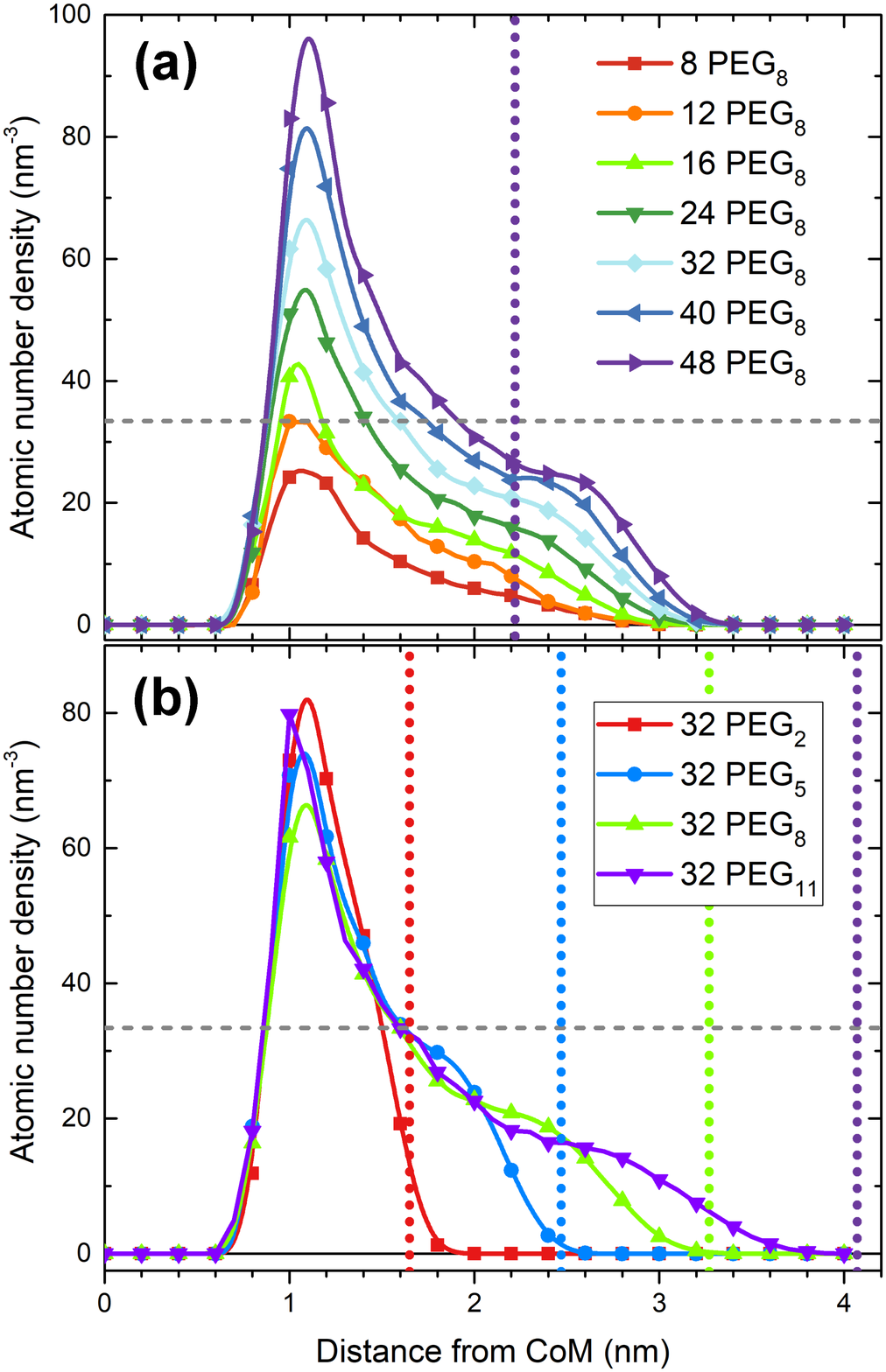}
    \caption{Atomic number density for the thiol-PEG-amine coatings in water as a function of distance from the CoM of the gold core.
    Panel~(a): number density distribution for thiol-PEG$_8$-amine for different number of ligands attached.
    Panel~(b): number density distribution for the coatings made of 32 ligand molecules with the different number of ethylene glycol monomers. Horizontal dashed lines show the number density of water molecules in ambient water, $n_{\rm wat} = 33.4$~nm$^{-3}$. The meaning of vertical dotted lines in panels~(a) and (b) is given in the main text. }
    \label{fig:PEG_density}
\end{figure}

In what follows we will focus on structural characterisation of \textit{solvated} systems which are relevant for experiments on NP radiosensitisation.
Figure~\ref{fig:PEG_density} shows the number density of all atoms in the selected thiol-PEG-amine coatings in water.
Figure~\ref{fig:PEG_density}(a) shows how the atomic number density changes upon varying the number of ligands attached to the metal core;
molecules containing eight ethylene glycol monomers (PEG$_8$) are considered as an example.
Figure~\ref{fig:PEG_density}(b) shows atomic number density for the coatings made of 32 molecules (corresponding to ligand surface density of 4~nm$^{-2}$) of different length.
The curves presented in Fig.~\ref{fig:PEG_density} and elsewhere in this section were averaged
over 40 independent frames taken from the final 2~ns long segments of the simulated trajectories.

Figure~\ref{fig:PEG_density}(a) shows that the shape of each number density curve
remains roughly the same for a given ligand length, independent of ligand count.
All number density curves peak at about 1.1~nm from the CoM, that is about 0.3~nm away from the surface of the metal core.
This peak is broader for small number of ligands attached while it becomes narrower and sharper as
the coating density increases (see Fig.~\ref{fig:PEG_density}(a)).
A broader shoulder centred at about 2.5~nm distance from the CoM emerges for denser coatings as shown
for the case of 48 thiol-PEG$_8$-amine molecules.
The formation of such a shoulder indicates that molecules in dense coatings of solvated NPs tend to unfold into elongated ``brush''-like structures which are seen also in Figure~\ref{fig:AuPEG_vac_vs_wat}(b).
This effect has a geometrical explanation and is related to the dense packing of ligands in the coating.
The vertical dotted line in Fig.~\ref{fig:PEG_density}(a) shows the thickness of a spherical shell which surrounds the gold core and
has a volume equal to the volume occupied by a single thiol-PEG$_8$-amine molecule in vacuum, multiplied by 48.
Due to electrostatic and van der Waals interactions between ligands and water molecules
such a dense packing of ligands is not possible and the ligands are stretched further from the metal core,
which results in an increase of atomic number density for coating molecules at large distance from CoM.
A more detailed analysis of the structure of ligands in solvated NPs is presented in the next section.

Note also that the maximal atomic number density for the coatings made of 16 or more thiol-PEG$_8$-amine molecules
is higher than the number density of water molecules in a liquid medium at ambient conditions, $n_{\rm wat} = 33.4$~nm$^{-3}$
(see a horizontal dashed line).
For dense coatings containing 40 molecules or more the number density for the coatings exceeds that of water
in about 50\% of the coating region.
As discussed below in
Section~\ref{sec:results_Water}
dense organic coatings hinder penetration of water molecules into the coating region.
As a result such coatings have low water content which may affect the radiosensitising properties of coated metal NPs.

\begin{figure}[t]
    \centering
    \includegraphics[width=0.85\linewidth]{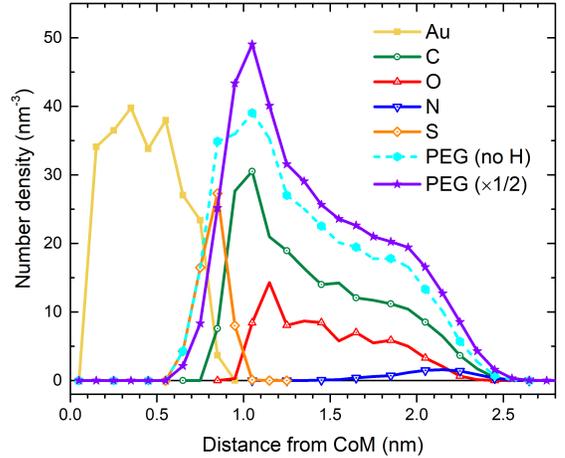}
    \caption{
    Number density of atoms of different type in the coating made of 48 thiol-PEG$_5$-amine molecules in water as a function of distance from the CoM of the gold core.}
    \label{fig:PEG_density_elements}
\end{figure}

\begin{figure*}[t]
    \centering
    \includegraphics[width=0.70\linewidth]{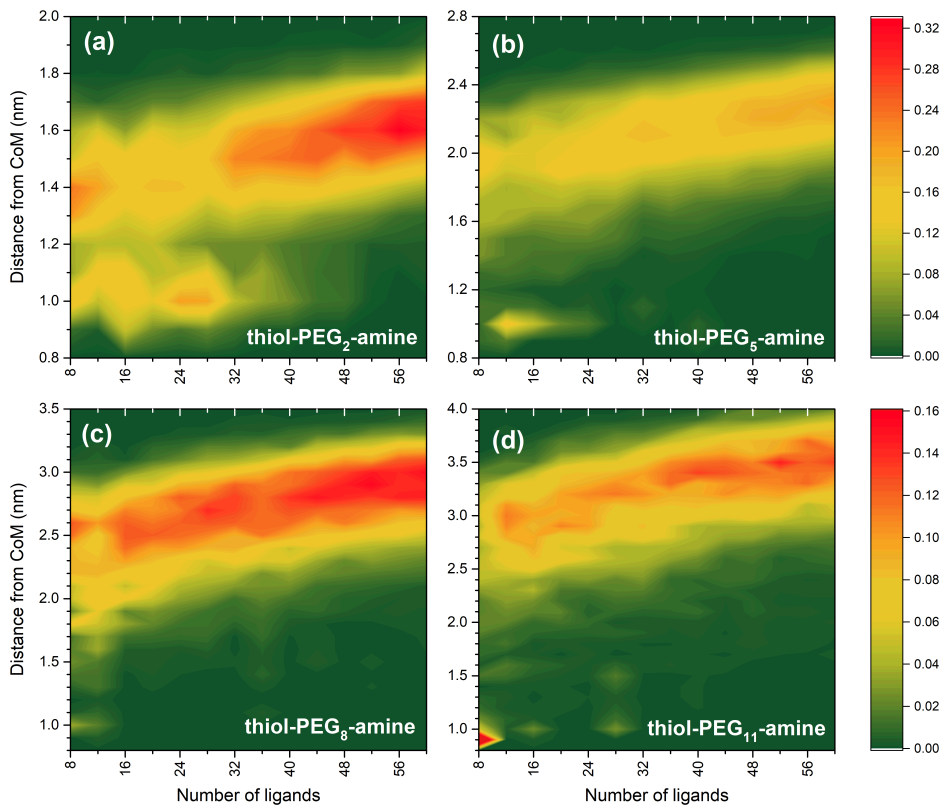}
    \caption{Probability distribution of nitrogen atoms in thiol-PEG$_n$-amine ($n = 2, 5, 8, 11$)
    coatings made of $8-60$ ligand molecules. Several distinct conformation states of ligand molecules are revealed at low-density coatings containing up to 24 ligands; this effect is especially pronounced for the coatings made of shorter PEG$_2$ and PEG$_5$ molecules. As both the number of ligands in the coatings and the ligand length increase the molecules acquire the elongated ``brush''-like shape. The color scale shows the probability of finding a nitrogen atom at a given distance from CoM.
    Note that the colour scale for thiol-PEG$_2$-amine and thiol-PEG$_5$-amine (panels~(a,b)) is twofold larger than for the longer ligands (panels~(c,d)).}
    \label{fig:nitrogen_contour}
\end{figure*}

Figure~\ref{fig:PEG_density}(b) shows that longer ligands in the coatings of a given surface density
are stretched further from the NP core.
The coating made of thiol-PEG$_{11}$-amine molecules extends up to about 4~nm from the CoM, that is 3.2~nm
from the surface of the metal core.
Note that maximum spatial extent of the thiol-PEG-amine coatings in water can be estimated without conducting time-consuming
multi-step calculations but just knowing the geometry of a single ligand molecule in vacuum (see Fig.~\ref{fig:Au135_Au_AuL}(b)).
The vertical dotted lines in Fig.~\ref{fig:PEG_density}(b) show the sums of the radius of the gold core and lengths of the optimised molecules in vacuum.
It is apparent that ligand molecules in relatively dense coatings made of 32 molecules have an elongated
shape that is similar to the shape of isolated molecules in vacuum.

The above-described analysis was performed for NPs which were first annealed in vacuum following
by solvation and equilibration in water.
A similar analysis was performed for the solvated systems produced using the alternative procedure, i.e.
when the NPs were solvated and then annealed together with the water medium.
We found that structures of those systems are very similar to the structures described above.
The final configuration of the systems is thus independent of which of the two procedures was used.

\begin{sloppypar}
Figure~\ref{fig:PEG_density_elements} breaks down the overall atomic number density of the coating into constituent elements,
allowing them to be characterised.
The coating made of 48 thiol-PEG$_5$-amine molecules (whose structure is shown in Fig.~\ref{fig:AuPEG_vac_vs_wat}(b))
is considered as an example.
Firstly, sulphur atoms (orange curve) are positioned at the surface of the NP core that indicates stability
of Au--S bonds and that no interlacing into the metal core had taken place during the annealing and
equilibration simulations.
The remaining elements are positioned in the same order as they appear in the ligand chain,
see Fig.~\ref{fig:Au135_Au_AuL}(a).
Nitrogen atoms (blue curve) are positioned at distances from 1.8 to 2.5~nm from CoM and do not appear
at smaller distances, which indicates that the ligand molecules are mostly positioned normally with
respect to the gold surface.
The sum of densities of all heavy elements (all elements except for hydrogen) is shown by a dashed blue curve.
The total number density including all atoms of the coating (divided by the factor of two) is shown by a purple
line with star symbols.
\end{sloppypar}

\subsection{Mushroom to Brush Transition with an Increase of Ligand Surface Density}

A detailed analysis of number density distributions for constituent elements permits us to elucidate
some common structural features of the thiol-PEG-amine coatings in an aqueous environment.

\begin{figure}[t]
    \centering
    \includegraphics[width=0.85\linewidth]{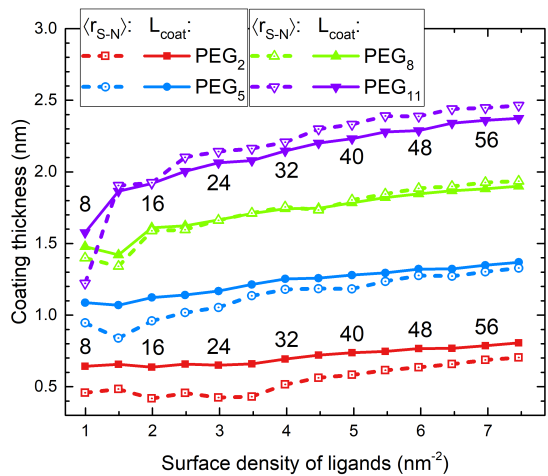}
    \caption{Coating thickness as a function of the surface density of ligands. The corresponding numbers of attached ligands are also indicated. Dashed lines with open symbols show the average distance between the sulphur and nitrogen atoms, $\left<r_{\rm S-N}\right>$. Solid lines with closed symbols show the thickness of the coating $L_{\rm coat}$ evaluated as the difference between the point containing 95\% of the coating atoms (including hydrogens) and the average position of sulphur atoms. }
    \label{fig:coating_thickness}
\end{figure}

Figure~\ref{fig:nitrogen_contour} presents the radial distribution of nitrogen atoms in coatings
with different lengths and surface densities of ligands.
The number of nitrogen atoms located in spherical bins of 0.1~nm thickness was calculated and
divided by the number of ligand molecules in a particular coating.
The total probability of finding a nitrogen atom integrated over distance from CoM is thus equal to unity.
As stated above the results presented were averaged over 40 independent simulation frames.
At ligand surface densities below 3 molecules per nm$^{2}$ (which correspond to up to 24 molecules attached)
the coating layer represents a mixture of different conformation states.
Nitrogen atoms in the low-density coatings are scattered over a wide range of distances from the CoM of the core;
this effect is especially seen for thiol-PEG$_2$-amine coatings.
The plot for the low-density coatings containing PEG$_5$ molecules also indicates a bimodal distribution
for the position of nitrogen atoms, whereas this effect becomes less prominent for the coatings made
of PEG$_{8}$ and PEG$_{11}$ molecules.
As shown in Figure~\ref{fig:nitrogen_contour} nitrogen atoms are preferentially distributed at larger distances from CoM,
whereas small islands corresponding to a higher probability of finding a nitrogen atom are also clearly seen
at about 1~nm distance from CoM.
The positions of nitrogen atoms are close to the calculated average position of sulphur atoms, 0.85~nm from CoM.
This indicates that low-density coatings made of shorter ligands have several distinct conformation states,
namely molecules curled tightly around the core as well as elongated ``brush''-like structures.
As both the surface density and the length of ligands increase the molecules acquire predominantly the ``brush''-like shape.
This behaviour is due to the fact that the characteristic distance from the CoM to nitrogen atoms increases gradually with increasing the density of ligands.

The formation of ``brush''-like structures is illustrated further in Fig.~\ref{fig:coating_thickness}.
Dashed lines with open symbols show the average distance between sulphur and nitrogen atoms, $\left<r_{\rm S-N}\right>$.
These values are compared with the estimated thickness of the coating in a liquid water medium,
$L_{\rm coat}$ (solid lines with closed symbols).
The latter is calculated as the distance between the point containing 95\% of the coating atoms
(including hydrogen atoms) and sulphur atoms.
For PEG$_2$ and PEG$_5$ coatings the estimated coating thickness significantly exceeds the average S--N distance,
which indicates that short ligand molecules are strongly bent.
For thiol-PEG$_8$-amine and thiol-PEG$_{11}$-amine coatings the average distance between sulphur and nitrogen atoms
becomes comparable or even larger than $L_{\rm coat}$.
This indicates that ligand molecules turn into more elongated ``brush''-like structures
as the coating becomes more dense.

\subsection{Distribution and Structural Properties of Water Around Coated NPs}
\label{sec:results_Water}

\begin{figure}[t]
    \centering
    \includegraphics[width=0.95\linewidth]{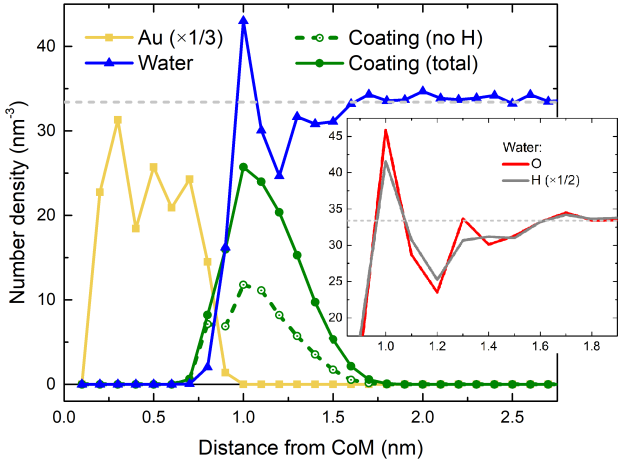}
    \caption{
    Number density distributions for gold atoms of the metal core (yellow line), all atoms of the coating made of 8~thiol-PEG$_2$-amine molecules (green lines) as well as for water molecules surrounding the NP (blue line).
    The inset breaks down the number density of water molecules into the contributions stemming from oxygen and hydrogen atoms.
    A strong peak in the number density of water molecules corresponding to the formation of a dense water layer in close proximity to the gold surface is clearly seen. }
    \label{fig:water_layer}
\end{figure}

We will now discuss an analysis of the water content and spatial distribution of water molecules in the vicinity
of coated metal NPs.
An increase of water content near the surface of the NP has been observed at low densities of the coating.
Figure~\ref{fig:water_layer} shows number density distributions for gold atoms of the metal core, all atoms of the
coating made of 8~thiol-PEG$_2$-amine molecules as well as for water molecules surrounding the NP.
A strong peak at 1~nm distance from CoM corresponding to an increase in the number density of water molecules
in close proximity to the gold surface is clearly seen.
A similar effect, namely a rearrangement of the first water solvation shell around nanometer-sized bare gold NPs, has been recently studied computationally by means of classical MD and quantum chemistry methods \cite{Tandiana_2021_JCP.154.044706}.

\begin{figure}[t]
    \centering
    \includegraphics[width=0.85\linewidth]{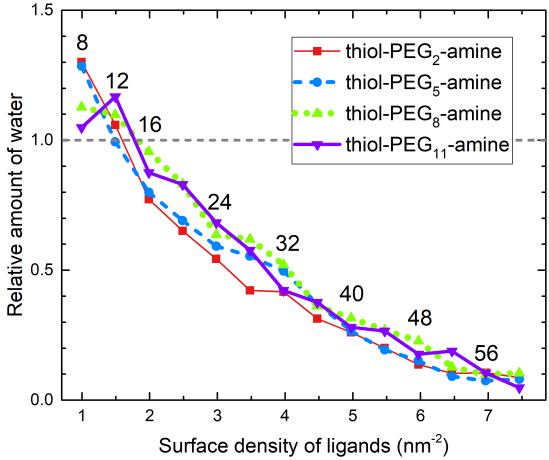}
    \caption{Relative change of the content of water around the gold NP as a function of ligand length and the surface density of ligands.
    The corresponding numbers of attached ligands are also indicated.
    The calculated number densities are normalised to the ambient number density of water (shown by the horizontal dashed line).
    An increase of the content of water by 10--30\% is seen at low-density ligands made of up to 12 molecules. }
    \label{fig:Water_increase}
\end{figure}

When the metal core is covered by a small number of ligands, a significant part of the metal surface is open
to surrounding water.
Due to strong van der Waals interaction between gold atoms and oxygen atoms of water molecules
the latter become attracted to the core and form a dense monomolecular water layer near the NP surface.
The maximum number density of water molecules in the thiol-PEG$_2$-amine coating is 43 nm$^{-3}$,
which is about 30\% higher than the ambient water density.

An increase of water content in close proximity to the metal core may have an important effect on
the production of hydroxyl radicals around the NP.
As shown previously \cite{Verkhovtsev_2015_PRL.114.063401, Verkhovtsev_2015_JPCC.119.11000} irradiation
of nanometre-sized metallic NPs with ions at Bragg peak energies induces the formation of plasmon excitations in the NPs.
Decay of these collective electron excitations results in a strong emission of low-energy electrons with the kinetic energy of a few eV.
Electrons of such energies can travel short distances (several Angstroms) from the surface of the NP \cite{haume2018transport}.
An increase in water density near the surface will thus increase the probability of electron interactions
with water molecules (e.g. dissociative electron attachment) and the subsequent formation of OH radicals.

Figure~\ref{fig:Water_increase} shows the number density of water molecules in a 0.1~nm-thick spherical shell
in proximity to the metal core for different coating densities and ligand lengths.
The calculated number densities are normalised to the ambient number density of water, $33.4$~nm$^{-3}$.
As shown in the figure, an increase of the amount of water has been observed for the low-density coatings
of different length (with the density up to 1.5 molecules per nm$^2$).

\begin{figure*}[t]
    \centering
    \includegraphics[width=0.75\linewidth]{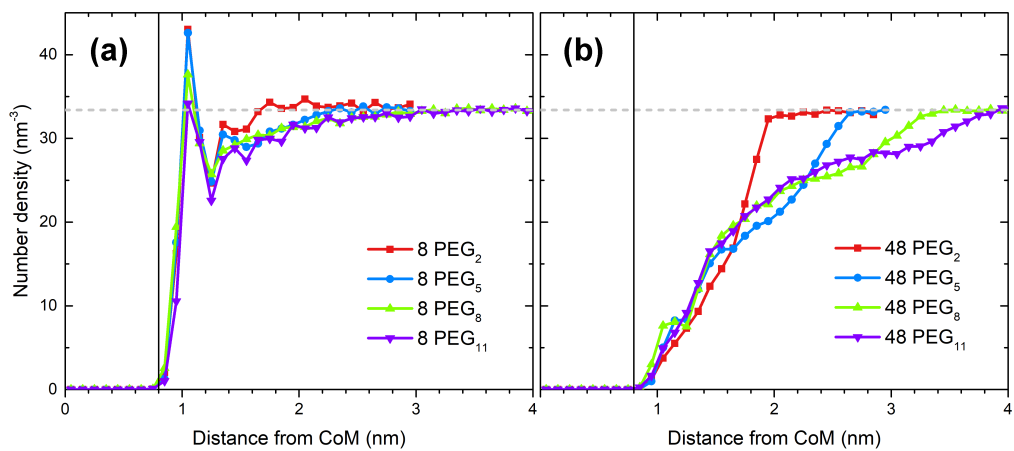}
    \caption{
    Number density of water molecules surrounding the gold NPs coated with 8 (panel~(a)) and 48 (panel~(b)) thiol-PEG$_n$-amine ligands of different length.
    The horizontal dashed line shows the ambient number density of liquid water.
    The vertical line denotes the position of the gold surface.
    }
    \label{fig:water_numdens}
\end{figure*}

The opposite effect has been observed for denser coatings where the amount of water near the NP surface
is significantly reduced.
When the ligand density is above 6 molecules per nm$^{2}$ the number of water molecules in the coating region
around the metal surface is five to ten times lower than the number of water molecules in ambient water.
Interestingly, this result is almost independent of the length of ligands for the molecules considered in this study.
The amount of water within the coating layer is not distributed uniformly but increases gradually with
distance from CoM, which is illustrated in Fig.~\ref{fig:water_numdens}.
However, for long thiol-PEG$_{11}$-amine coatings it is still lower than the ambient water density at
distances up to 3~nm from the gold surface.
A reduced amount of water in close proximity to the metal core will suppress the production of OH radicals
by low-energy electrons emitted from the metal NP.
High-density coatings (even made of short molecules like thiol-PEG$_2$-amine) may thus impair
the radiosensitising properties of metal NPs with regard to this molecular mechanism of radiosensitisation.

\section{Conclusion}

We presented a detailed atomistic approach towards computational modelling and structural characterisation
of experimentally relevant coated metal nanoparticles (NPs) in explicit molecular environments.
The understanding of NP structure (especially that of the coating) and interaction with a solvent is crucial for determining the potential efficiency of coated metal NPs as radiosensitisers in radiotherapy applications.
The computational procedure laid out in this paper can be utilised for many other combinations
of core and coating, in a number of environments (e.g. given temperature, composition and density of solvent).
The benefit of using computer simulations is that many such systems can be quickly generated and analysed
in comparison to costly experiments.

\begin{sloppypar}
As an illustrative and experimentally relevant case study we systematically analysed the structural properties
of 1.6~nm diameter gold NPs coated with thiol-PEG$_n$-amine molecules of different length ($n = 2,5,8$ and 11).
The surface density of ligands was varied from 1 to 7.5 molecules per nm$^{2}$ for each ligand length.
\end{sloppypar}

We found that structure of the coating layer of the solvated NPs depends strongly on the surface density of ligands.
At densities below about 3 molecules per nm$^{2}$ the coating represents a mixture of different conformation states,
whereas elongated ``brush''-like structures are predominantly formed at higher surface densities.

Simulations revealed that a dense water layer is formed near the surface of the NP at low densities of the coating.
This effect occurs due to strong van der Waals interaction between gold atoms and oxygen atoms of water molecules.
A 10--30\% increase of water density at a few Angstroms distance from the gold surface may enhance the production
of hydroxyl radical due to low-energy electrons emitted from metallic NPs \cite{Verkhovtsev_2015_PRL.114.063401, Verkhovtsev_2015_JPCC.119.11000, haume2018transport}.
Therefore, low-density coated NPs may possess enhanced radiosensitising properties as compared to their more densely
coated counterparts.
For denser coatings, the density of water in the coating layer is lower than the ambient water density at distances
from 1~nm up to 3~nm from the gold surface depending on the length of ligand molecules.
A low water content near the metal core should decrease the probability of radical production and thus diminish
the radiosensitising properties of such NPs.

The research reported in this paper can be extended in several possible directions.
Firstly, the effects arising upon irradiation of such coated metal NPs can be studied with the aim to understand
how the atomistic structure of the coating (including the amount and spatial distribution of water within the coating layer)
affects the transport of secondary electrons and the production of free radicals around the NPs \cite{haume2018transport}.
Beyond that, a range of metal cores of different composition and size (including highly symmetric ligand-protected gold clusters \cite{pohjolainen2016unified, Jin_2010_Nanoscale.2.343, Wang_2021_NanoscaleAdv.3.2710}), coatings and environmental conditions could be explored in a more systematic manner and compared with experimental observations.

\section*{Acknowledgments}

This work has received financial support from the European Union's Horizon 2020 research and innovation programme
-- the Radio-NP project (Grant agreement no. 794733) within the H2020-MSCA-IF-2017 call and
the RADON project (Grant agreement no. 872494) within the H2020-MSCA-RISE-2019 call.
The work was also supported in part by the COST Action CA20129 ``Multiscale Irradiation and Chemistry Driven Processes and Related Technologies'' (MultIChem).


\bibliography{bib}

\end{document}